# The Gendered Cost of Lower Grades: Women's Physics Perceived Recognition and Identity Suffer Disproportionately If They Earn Less Than A Grade

Jaya Shivangani Kashyap[1*], Christian D. Schunn[2] and Chandralekha Singh[1]

[1] Department of Physics and Astronomy, University of Pittsburgh, Pittsburgh, PA 15260

[2]Learning Research and Development Center and Department of Psychology, University of Pittsburgh, Pittsburgh, PA 15260

*jak509@pitt.edu

## Abstract

Perceptions of disciplinary recognition and identity can be shaped by various forms of feedback and experiences. Here we focus on the potential effects of course grades on the perceievd recognition and physics identity of students. We analyze patterns in changes in physics identity and perceived recognition from pre course to post course across three cohorts of university students enrolled in calculus-based Physics 1 ($N = 1{,}681$). Students not receiving A grade, on average, showed declines in physics identity and perceived recognition. Even a B grade resulted in declines, and the declines were nonlinear across lower grades. Changes in perceived recognition fully mediated the changes in identity. Importantly, women showed significantly larger declines in identity and perceived recognition, compared to men, if they got less than A grade. The gender moderation was specifically localized to changes in perceived recognition, with no further gender effects on identity beyond the cascading effects on perceived recognition.

## Introduction

There have been many efforts to improve the participation and retention of women in science, technology, engineering, and math (STEM) degrees (Walton et al., 2015; White et al., 2021), including changes in pedagogical methods (Beach et al., 2012; Stolk et al., 2021), representations of women among instructors and in textbooks (Good et al., 2010; Kostas, 2021), and psychosocial interventions seeking to address classroom climate concerns (Walton et al., 2015). Unfortunately, overall progress in increasing the number of women in physics and other STEM domains such as computer science, electrical engineering, and mathematics is very slow, and significantly slower than in all other natural science disciplines and most areas of STEM (Ceci et al., 2014; Cheryan & Plaut, 2010; Hill et al., 2010; Maltese & Cooper, 2017; Means et al., 2018). While some interventions show promise (Clark et al., 2015; Wulff et al., 2018), overall, more research is needed to better understand all factors that shape participation in physics.

Recently, education researchers have placed growing emphasis on students' beliefs about themselves in relation to STEM that are associated with performance in STEM (Binning et al., 2020; Broda et al., 2018; Marshman, Kalender, Nokes-Malach, et al., 2018; Theobald et al., 2020). Key beliefs that have drawn attention include students' disciplinary self-efficacy, interest, perceived recognition, and identity (Maries et al., 2021). In addition, interventions explicitly addressing these beliefs have been shown to improve student learning and reduce achievement gaps (Binning et al., 2020).

In addition, these beliefs can also serve as key mediators in the impact of the environment (e.g., cultural stereotypes, early science learning experiences in and out of school, role models) on participation (Beasley & Fischer, 2012; Chemers et al., 2011; Cimpian et al., 2016; K. Whitcomb et al., 2021). In particular, career choices and persistence are strongly associated with one's disciplinary identity (Hazari et al., 2010; Jiang et al., 2025). For example, in the case of physics, career choices for physics and physics-related majors are predicted by having a physics identity (R. M. Lock et al., 2013).

While prior research has established that identity and perceived recognition are important predictors of persistence (Casserly & Rock, 1985) in physics and physics-related majors, there is still limited understanding of how these beliefs change over the course of a semester and what factors influence those changes. A salient factor for students in general is the grades they receive in a course (Bottomley et al., 2022; Marshman, Kalender, Nokes-Malach, et al., 2018). However, the relationship between different course grades and changes in identity and recognition remains poorly understood. Ideally, students respond to moderately lower performance in just one class by changing their study strategies rather than by giving up on their career pathways. However, students may take such early performance feedback as identity-shaping evidence. Furthermore, it remains unclear how these patterns differ across men and women, although there is prior work suggesting that women react more strongly to negative performance feedback in the context of a discipline that is male-dominated (Marshman, Kalender, Nokes-Malach, et al., 2018). This study addresses these gaps by examining how students' physics identity and perceived recognition change over time, how these shifts relate to course grades and gender.

## Literature Review

Many formal models of disciplinary identity have been explored, including those focused on how identity shapes interactions (Gee, 2000) and the beliefs such as interest and self-efficacy (Gee, 2000; Smith, 2012) that underlie identity formation. Individuals can have multiple identities at the same time (Gee, 2000; Stets & Burke, 2003), and which identities are activated can vary by context. For example, one may have a different identity within home and school contexts (Gee, 2000). For students to continue along a formal education pathway, it is likely important that they hold academic/discipline-based identities in formal contexts.

At its core, identity involves being recognized as a certain 'kind of person' (Gee, 2000), such as being a physics kind of person. Across theoretical frameworks, there has been some debate about the internal (how a person feels about themselves) vs. external (how others feel about them) nature of identity (Rodriguez et al., 2019; Stets & Burke, 2003). For example, whether a person manifests themselves as a physics person can depend upon how they act and feel in a physics classroom. Regardless of whether the formal label of identity is applied to only external aspects, only internal aspects, or some combination of the two, a common assumption is that how a person feels about their own identity is shaped by their perceptions of how others feel about them (Carlone & Johnson, 2007; Hazari et al., 2010; Rattan et al., 2012; Wigfield & Eccles, 2002; Zimmerman, 2000). That is, students are more likely to think they are a physics kind of person if important people around them (e.g., friends, family, instructors) recognize them as a physics kind of person (Hazari et al., 2010; Kalender et al., 2019a).

Identity is viewed as a critical concept in STEM research because it appears to be an especially strong predictor of continued participation in STEM contexts. In middle and high school students, having a science identity is an especially strong predictor of whether students wish to enroll in optional advanced science courses as well as engage with a wide variety of informal science learning opportunities, such as after-school programs, clubs, and summer camps (Vincent-Ruz & Schunn, 2018), even when controlling for self-efficacy and interest. Having a science identity in early university levels predicts continued enrollment into larger science coursework (Robinson et al., 2019) and wanting to pursue careers related to the specific science disciplinary identity (e.g., a physics identity predicting physics-related careers; (Hazari et al., 2010)). A recent meta-analysis of 35 studies revealed that STEM identity and STEM career intention showed a strong relationship on average of $r = .35$, although with some variation across contexts and student groups (Jiang et al., 2025).

Turning to the central topic of representation, identity, and gender, large differences in physics identity by gender have been observed at the beginning of introductory physics courses for natural science and engineering majors (Kalender et al., 2019a, 2019b; Li et al., 2020a), as well as such courses for majors in biology and health science pathways (Cwik & Singh, 2021a, 2021b, 2022b). Throughout schooling, women are less likely to be recognized as physics kind of people given cultural stereotypes (Hazari et al., 2007; Maries et al., 2025), which then shapes their many physics experiences (e.g., enrolled in optional or advanced physics courses, attended physics-related summer or after-school programs) and positive physics experiences prior to attending university. High school (Aschbacher et al., 2010) plays an especially important role in developing the physics identity of women (R. M. Lock et al., 2015). For example, recognition from physics teachers has been identified as playing an important role in students choosing physics as a career (Hazari & Cass, 2018).

Although university instructors have little control over the factors that shape student identities (and other psychosocial beliefs more generally) before attending university, they can play a role in reversing these initial differences (Mueller & Dweck, 1998). Unfortunately, beliefs about physics for women tend to decline across the sequence of introductory courses(T. Nokes-Malach et al., 2017) (Van Dusen, 2025), including physics identity (Kalender et al., 2019a; R. M. Lock et al., 2013). Women bring many assets that lead them to perform at slightly higher levels in many types of courses in high school and university (Voyer & Voyer, 2014), including in mathematics (Voyer & Voyer, 2014), which is a key performance asset for physics given its intensely mathematical nature. However, physicists are mostly portrayed as men who are also considered to inherit a natural ability to be successful because of being a "genius" and "brilliant" (Li et al., 2020b).

These differential experiences and assets raise the question of what kinds of experiences within a university course shape students' physics identity and whether these experiences affect men and women in a similar manner. A common assumption is that students' identity is shaped by their performance (Carlone & Johnson, 2007). That is, when students do well in physics (e.g., on an exam, in a course, at a competition), they may be more likely to think of themselves as a physics person (Hazari et al., 2010). To capture these effects in theoretical models, the science identity framework developed by Carlone and Johnson (2007) includes three interrelated components that shape identity: competence, performance, and recognition. Hazari et al. (2010) adapted this framework specifically for the context of physics. In their adaptation, competence was reframed as understanding physics concepts, while performance was defined as the ability to carry out physics-related tasks (e.g., doing well in exams or in other course-related tasks). They defined recognition as the extent to which students feel recognized by others as being good at physics, and added a fourth dimension, interest, as a factor that shaped identity. Taking a slightly different take, Kalendar and colleagues conceptualized physics identity as being composed of only beliefs: perceived recognition, interest, and self-efficacy (Kalender et al., 2019a, 2019b), choosing not to include external factors in the internal construction of identity. But conceptually, the self-efficacy component would be influenced by performance, so the theoretical difference is small. Variations of this model have been validated and refined through a mixture of quantitative (Avraamidou, 2021; Cribbs et al., 2015; Danielsson, 2009; Dou & Cian, 2022; Verdín et al., 2018) and qualitative data (Avraamidou, 2021; Bottomley et al., 2022; Li & Singh, 2023). Overall, there is consistent theoretical support for performance as a key factor shaping identity, though there may be disagreement over whether its effects on identity are direct or mediated by its influence on other constructs, such as perceived recognition.

Further, the extent to which grades in particular have a linear effect on identity is an open question. Pragmatically, when students receive failing grades (or very low passing grades, which are not sufficient to enroll in the next course), their ability to continue on a career pathway is objectively in peril. Statistically, receiving a low grade in an early course may predict receiving non-passing grades in future, more difficult courses or failing to progress into advanced degrees (e.g., having a sufficiently high GPA to enter a master's or PhD program). However, receiving a mid-level grade (e.g., a B) is a common experience in challenging STEM courses (K. M. Whitcomb & Singh, 2020) and should not be taken as a negative indicator of long-term success.

In addition, relatively little research has examined whether performance matters to the same degree for different student groups. Looking beyond research on identity, men and women tend to react differently to the same performance levels, particularly in domains like physics for which there are strong gender stereotypes (Marchand & Taasoobshirazi, 2013; Maries et al., 2022; Matz et al., 2017; Steele & Aronson, 1995). For example, one study found that men receiving Cs in physics had the same physics self-efficacy as women receiving As (Marshman, Kalender, Nokes-Malach, et al., 2018).

It is therefore an open question whether grade feedback shapes student identities similarly for women and men. On the one hand, given the different relationship of grades to self-efficacy, women may interpret lower grades as a stronger threat to their identity than do men. On the other hand, those receiving lower grades might already have lower physics identity, and no further negative effects are observed from this latest performance feedback. If there is a gendered reaction to the same grades, then changing how grade feedback is delivered may be needed to shape its gendered effects on students' beliefs.

Another open question concerns the impact of grades on internal identity versus perceived recognition. On the one hand, since grades are given to students privately (in the US), the effects may primarily influence internal identity. On the other hand, students do interact with instructors and TAs who are aware of their performance, and they can disclose their performance to friends and family. The reactions of others may be gendered (Danielsson, 2012; Day et al., 2016; Moss-Racusin et al., 2012), which could then produce gendered perceptions of being recognized as a physics person by those individuals. Thus, there may be especially large gender differences in perceptions of recognition as a physics person at different grade performance levels for some or all of these important external sources of recognition.

## Research Questions

Overall, the literature suggests that performance indicators such as grade are likely to influence identity and that perceived recognition is often a key precursor for identity. However, there are open questions about the specific nature of the relationship and how it is gendered. Therefore, with respect to relationships of physics course grade to changes in physics identity and perceived recognition, we ask three specific research questions:

RQ1: What is the form of the relationship between course grade and changes in both constructs?

RQ2: Do changes in perceived recognition mediate the relationship between course grade and changes in identity?

RQ3: Does gender moderate the relationships between course grade and changes in identity overall and at particular points in the mediated relationship?

We hypothesize that students feel less recognized as a physics person the lower the grade they receive, which then influences their internal physics identity. We further hypothesize that these effects are larger in women than in men, particularly in the extent to which lower grades shape being perceived as less of a physics person. Figure 1 shows the proposed model for the relationship between course grade of students and their identity, in which the effects of course grade are mediated through perceived recognition, and the mediation pathway is moderated by gender.

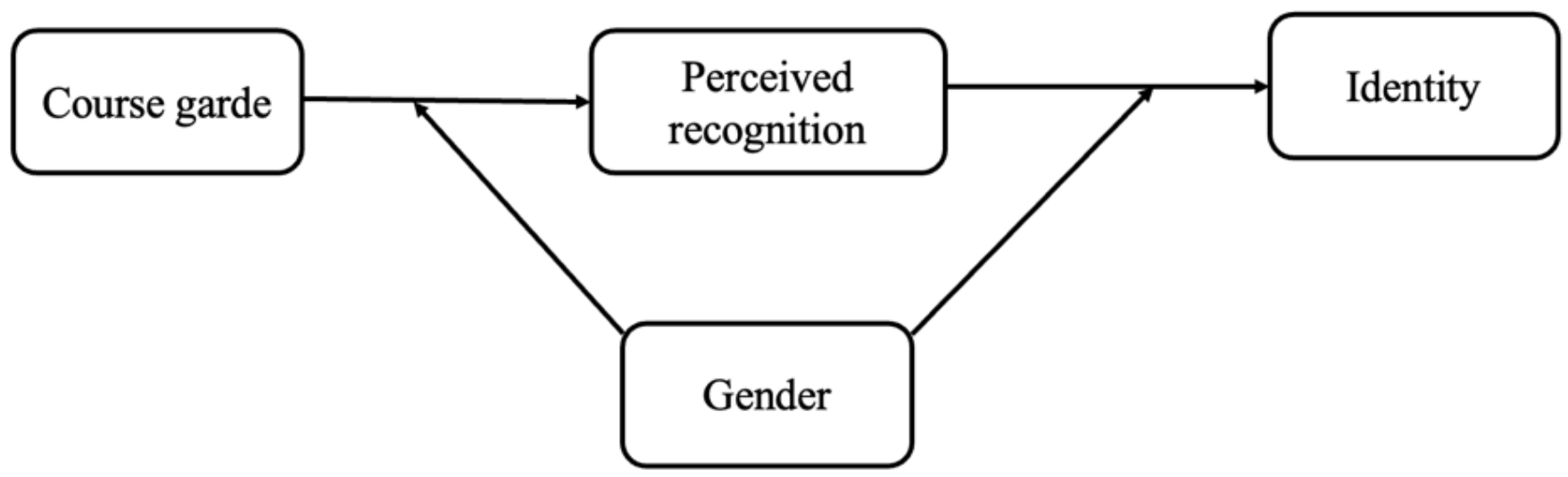

*Figure 1. Hypothesized model for the gender-moderated relationship between students' course grade and their identity through perceived recognition.*

## Methods

### Participants

This study included all students enrolled in Fall term sections of Calculus-based Physics 1 across three consecutive years at a large public research university in the US. These years were after the period in which instruction was heavily online due to the pandemic (i.e., class instruction was conducted in person, and students could interact normally with one another in class). Most students in this course are physical science (i.e., physics and chemistry) or engineering majors for whom this course is mandatory. Among the engineering majors, they were first-year students who had not yet selected more specific majors, including those closely related to physics (e.g., mechanical and electrical engineering) and those more closely related to other sciences (e.g., chemical, industrial, and bioengineering). Physics 1 includes the main content from mechanics, such as kinematics, Newton's laws of motion, energy and momentum, rotational kinematics and dynamics, gravitation, simple harmonic motion, and waves.

Student gender was provided by the university through an honest broker process that linked gender, grades, and survey data via an anonymized ID. A total of 64 students reported gender other than woman or man, which were too few to be included in meaningful statistical analysis. The remaining 1,681 students over the three years included were approximately 1,034 (62%) men and 647 (38%) women. This gender distribution is typical of physical science and engineering majors overall in the US (NSF, 2019).

**Measures**

*Identity and Perceived Recognition*. Students in this course regularly completed a longer survey involving a number of beliefs that were previously validated via extensive psychometric analyses and cognitive interviews with a range of students to ensure the items were being interpreted as intended in this context (Cwik & Singh, 2022b). The survey was given in the first two weeks at the beginning of the semester as a pre-survey and in the last two weeks of the semester as a post survey. In this study, we focus on two beliefs: 1) Physics identity, which is measured using single question about the extent to which students see themselves as a physics person (*I see myself as a physics person*); and 2) Perceived recognition in the context of physics, is measured using four items involving the extent to which students think other people (i.e., family, friends, TA, and instructor) see them as a physics person (e.g., My family sees me as a physics person) (Hazari et al., 2013; Hazari et al., 2010; Kalender et al., 2019a, 2019b). Note that single-item disciplinary identity measures have been found to be reliable and valid (Wu et al., 2024).

Across the three years of this study, the survey scale format changed after the first year. In the first year, the survey questions were presented on a 4-point Likert scale: strongly disagree, disagree, agree, and strongly agree. To provide opportunity for more nuanced responses, the scale was switched to a 7-point scale in the second and third years: strongly disagree, disagree, slightly disagree, neutral, slightly agree, agree, and strongly agree. Appendix present analyses showing high consistency in responses among students who completed the surveys in both formats only one month apart, in both absolute terms and also relative to the level of stability observed when the response format was held constant.

For perceived recognition, each student's score was calculated as the average across the four perceived recognition items. Cronbach's $\alpha$ (Cronbach, 1951) was used to assess the internal consistency across these items, with values above 0.8 generally indicating good reliability. $\alpha$ values were 0.86 and 0.92 for pre- and post-course perceived recognition, respectively, demonstrating strong internal consistency.

*Course grades.* Final course grades for Physics 1 were obtained and grouped into four categories: A (A+, A, or A-), B (B+, B, B-), Passing C (C+, C), and Non-passing (C-, D+, D, F). Grade groups rather than raw grades were used to support analysis of possible nonlinear effects and grade by gender interactions with sufficient statistical power (see Table 1).

*Table 1. Number of men and women in each grade group category, and relative percentages within each gender.*

| Gender | Grade Category | | | |
|---|---|---|---|---|
| | **Non-Passing** | **Passing C** | **B** | **A** |
| Men | 144 (14%) | 255 (25%) | 360 (35%) | 275 (27%) |
| Women | 112 (17%) | 192 (30%) | 211 (33%) | 132 (20%) |

**Analysis**

A subset of the 1,681 students enrolled in the studied classes completed the survey at pre- ($n = 1{,}472$) and at post- ($n = 1{,}116$). Since student scores were not missing at random (e.g., students with higher grades were more likely to complete the surveys), it was important to ensure that the results were not biased by systematic attrition. We therefore imputed missing student scores using multiple imputation by chained equations, as implemented in the MICE package in R (version 4.4.1). We compared the analysis of results using imputed scores with those obtained using only non-missing matched cases (i.e., keeping only those students who were present in both the pretest and posttest). We found qualitatively similar results and trends. Since the number of students missing from the non-passing grade group was high and there were not many students to use for imputation, the error bars were quite large with the imputed scores. Therefore, we present the analysis of results involving the non-imputed matched student scores.

The descriptive analysis of the identity and perceived recognition for men and women is calculated by finding means and standard errors. The effect sizes between pre and post identity and perceived recognition for men and women are calculated using Cohen's *d,* which is given as, $d = \sqrt{\frac{(n_1-1)*SD_1^2+(n_2-1)*SD_2^2}{n_1+n_2-2}}$. An effect size is typically considered small, medium, and large for values around 0.20, 0.50, and 0.80, respectively (Cohen, 1988). $n_1$ and $n_2$ represent the number of students in each group, $SD_1$ and $SD_2$ represent the standard deviation of those groups, respectively.

To answer the research questions, we performed linear regressions using the *regress* command and mediation and moderated mediation analysis using bootstrap in STATA version 17.0. In particular, the analyses involved either post-identity or post-perceived recognition as the outcome variables. For the first research question, we included grade group as a predictor, with dummy codes for each group in the regression analysis. The A group was used (here and in all analyses) as the reference category (e.g., interpreting the effect of a B in terms of the changes in identity after receiving a B rather than an A). Predicted post means adjusting for pre levels in identity or perceived recognition for each grade group were then calculated using the *margins* command. This method (a linear regression model predicting post, controlling for pre) is preferred over change-score measures because the latter tends to produce regression-to-the-mean artifacts, particularly with scales based upon few items.

For the second research question, we analyzed the mediation of the relationship between grade group and identity via perceived recognition. Similar to the regression methods for the first research question, post of each measure included a control for the corresponding pre measure so that changes in beliefs are being studied.

For the third research question, we analyzed the moderation of the overall relationship between grade and identity and perceived recognition, again controlling for pre as we did for the analyses in the first research question. Finding that there was significant moderation for both (and that perceived recognition mediated the relationship between grades and identity), we tested which parts of the mediation model developed in the second research question were significantly moderated by gender (see Figure 2).

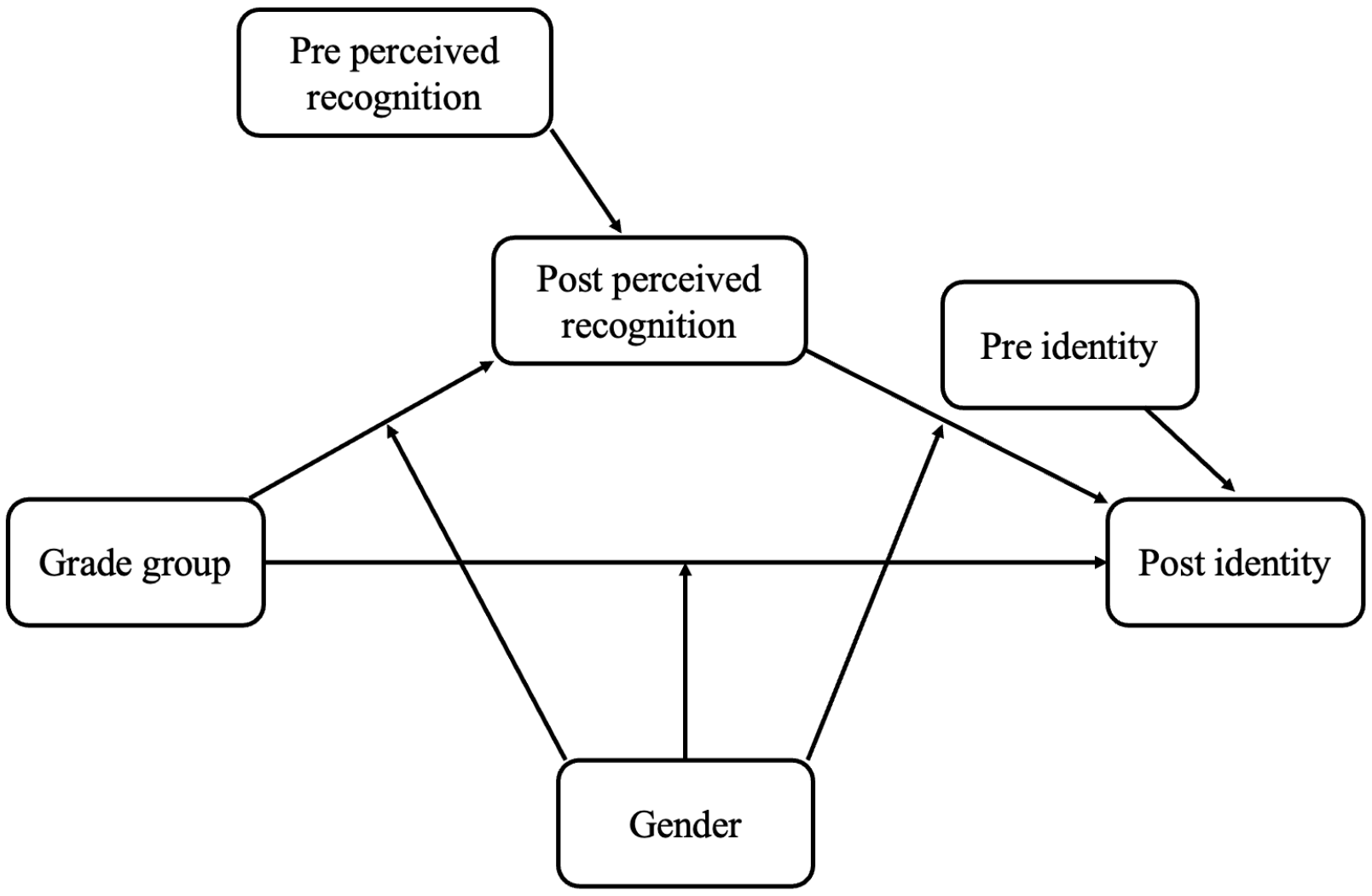

*Figure 2. The tested moderated mediation model, where gender potentially moderates either step of the mediation of grade to post identity through perceived recognition or the direct connection from grade group to identity.*

## Results

The mean values, number of students, and standard errors for pre-identity, post-identity, pre-perceived recognition, and post-perceived recognition for men, women, and all students are presented in Table 2. The effect sizes (Cohen's *d*) between pre- and post-identity for men and women are -0.06 and -0.28, respectively. Effect sizes between pre- and post-perceived recognition for men and women are 0.11 and -0.23, respectively. Even when grades are ignored, the effect sizes are small, but gender differences in identity and perceived recognition begin to emerge.

*Table 2. Mean, standard error, and number of students, pre and post, for identity and perceived recognition, for men, women, and all students.*

| | | Identity | | Perceived recognition | |
|---|---|---|---|---|---|
| | | Pre | Post | Pre | Post |
| Men | Mean (SE) | .22 (.02) | .19 (.02) | .07 (.01) | .11 (.02) |
| | *N* | 933 | 642 | 907 | 636 |
| Women | Mean (SE) | .02 (.02) | -.13 (.03) | -.04 (.02) | -.14 (.02) |
| | *N* | 580 | 485 | 565 | 480 |
| All | Mean (SE) | .15 (.01) | .05 (.02) | .02 (.01) | .00 (.01) |
| | *N* | 1,513 | 1,127 | 1,472 | 1,116 |

**RQ1: What is the form of the relationship between course grade and changes in identity and perceived recognition?**

*Identity.* Grades had a small positive relationship with pre-identity levels (see Figure 3a), making it clear why an investigation of the relationship of post-identity to grades would need to control for pre-identity levels. Also, mean levels were generally positive, meaning that on average, students most typically had at least a mild-positive identification with physics regardless of the grades they would eventually receive.

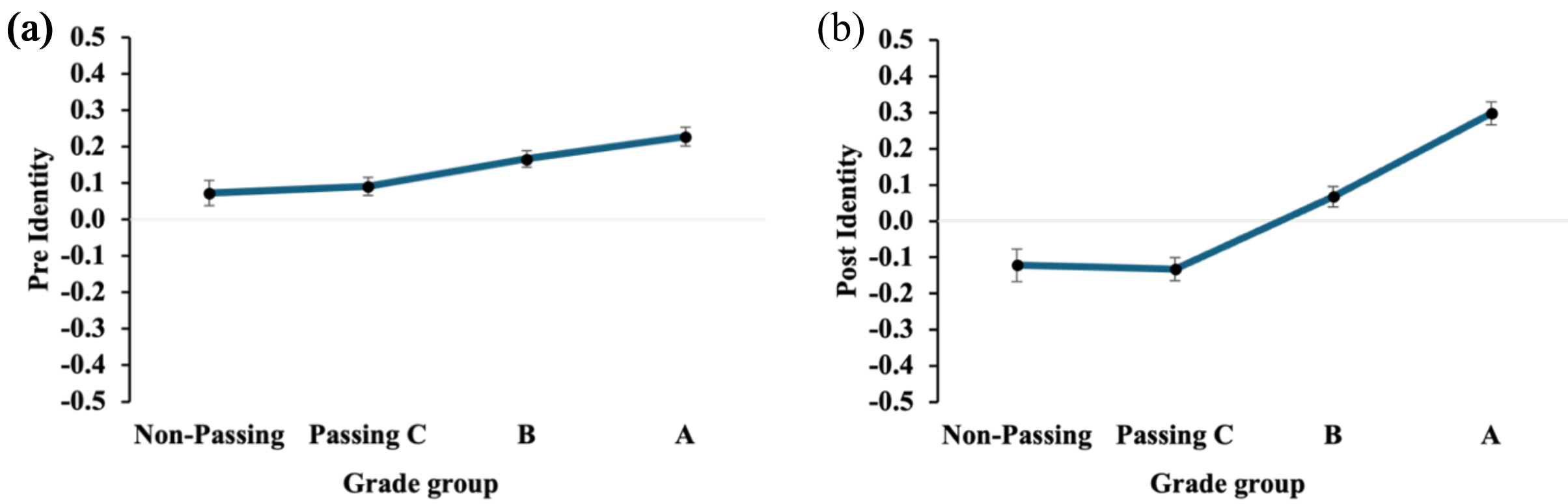

*Figure 3. Means (with SE bars) for (a) pre identity and (b) post identity (controlling for pre identity) across different grade groups.*

The regression model predicting post identity using pre identity and grade group accounted for approximately 40% of the variance in post identity. Post-identity values were generally well predicted by pre-identity values, as would be expected given the generally high pre-post correlation (see Table 3). Importantly, relative to receiving an A, receiving any other grade was associated with a significant decline in post-identity values, with large declines observed for non-passing grades and for passing C grades.

*Table 3. Regression coefficients for post identity in relationship to pre identity and grade group, with standardized $\beta$, t, and p-values.*

| Predictor | Standardized $\beta$ | $t$ | $p$ |
|---|---|---|---|
| Pre-identity | 0.54 | 21.36 | <0.001 |
| Non-passing | -0.20 | -7.31 | <0.001 |
| Passing C | -0.30 | -9.89 | <0.001 |
| B | -0.17 | -5.42 | <0.001 |

Figure 3b presents the predicted means for post-identity for each grade group, controlling for pre-identity. The post-identity means for each grade group, without controlling for pre-identity, are shown in Figure A1 in the Appendix. Students with A grades tended to continue to identify themselves as a physics person at the end of the course at similar or slightly stronger levels. On the other hand, there were sizeable declines in identity for those receiving a B, although the mean levels remained positive. By contrast, the declines were much larger for those receiving either a passing C or non-passing grade, with students now tending to have negative perceptions of their physics identity at the end of the course.

*Perceived recognition.* Similar to identity, student's pre perceived recognition levels had a small positive relationship to the grades they would eventually receive (see Figure 4a). Overall, students tended to be close to the neutral point on the perceived recognition scale except for those receiving an A.

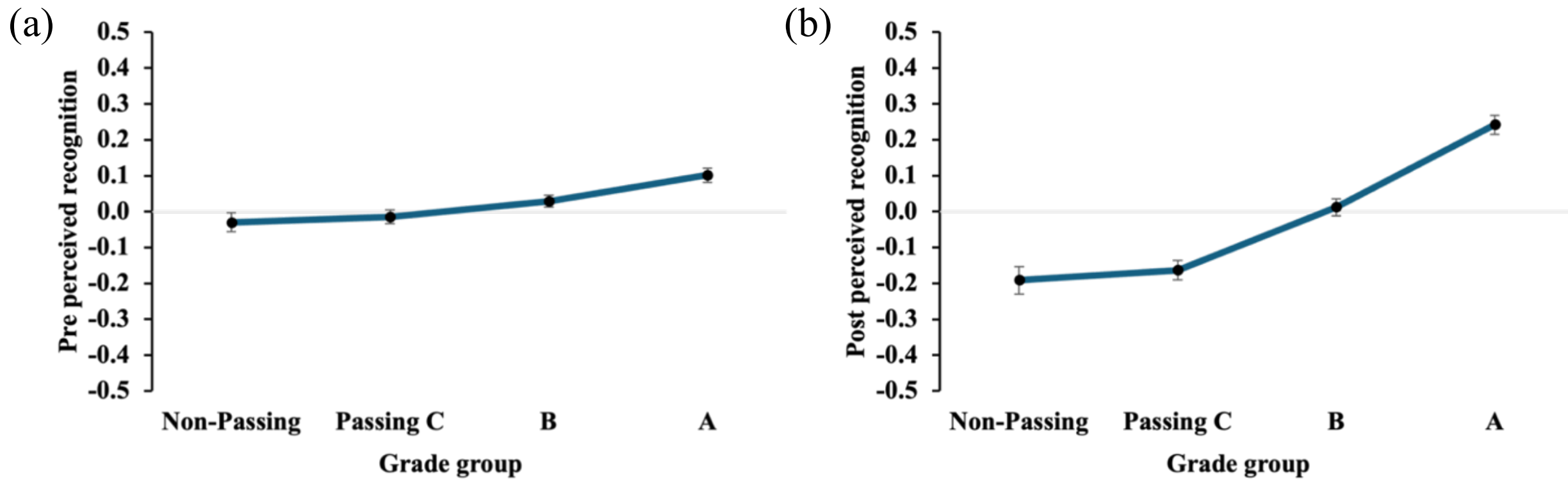

*Figure 4. Means (with SE bars) for (a) pre-perceived recognition and (b) post-perceived recognition (controlling for pre-perceived recognition) across different grade groups.*

Similar to identity, post-perceived recognition values were also well predicted by pre-perceived recognition, resulting from high pre-post correlation in perceived recognition. However, there were also significant changes in students' perceived recognition at the end of the course, which were well predicted by the grades they received (see Table 4). Figure 4b presents the predicted means for post-perceived recognition for each grade group, controlling for pre-perceived recognition. This model accounted for 43% of the variance in the post-perceived recognition. Means for post-perceived recognition for each grade group, without controlling for pre-perceived recognition, are shown in Figure A2 in the Appendix.

*Table 4. Regression coefficients for post-perceived recognition in relationship to pre-perceived recognition and grade group, with standardized β, t and p-values.*

| Predictor | Standardized $\beta$ | $t$ | $p$ |
|---|---|---|---|
| Pre perceived recognition | 0.56 | 22.13 | <0.001 |
| Non-passing | -0.24 | -8.49 | <0.001 |
| Passing C | -0.29 | -9.74 | <0.001 |
| B | -0.17 | -5.51 | <0.001 |

Similar to identity, all grades below an A were associated with declines in perceived recognition. Further, those with an A tended to have more positive mean levels of perceived recognition at post than at pre. By contrast, those with a passing C or a non-passing grade finished with negative perceived recognition levels on average. In other words, changes in both identity and perceived recognition were strongly related to grades.

**RQ2: Do changes in perceived recognition mediate the relationship between course grade and identity?**

The mediation model results (see Table 5) strongly support the mediation hypothesis. Post-perceived recognition is a strong predictor of changes in identity (i.e., the direct connection to post-identity when controlling for pre-identity is large and highly significant). Further, in contrast to the significant effects of receiving anything other than an A on changes in perceived recognition, the direct effects of grades on changes in identity are non-significant, including even the two kinds of C. Furthermore, this difference in what is statistically significant is not a matter of low power;

the 95% CIs for grade effects on perceived recognition are very different from the 95% CIs for grade effects on post-identity after including the perceived recognition mediator. The relationship between course grade and post-perceived recognition is therefore mediated by post-perceived recognition.

*Table 5. Unstandardized β, bootstrap SE, z, p, and confidence interval for the mediation model in which grade group and post-identity are mediated by post-perceived recognition, controlling for pre-identity and pre-perceived recognition. Statistically significant βs are bolded.*

| | Unstandardized *β* | Bootstrap *SE* | *z* | *p* | 95% CI | |
|---|---|---|---|---|---|---|
| *Predicting Post-Perceived Recognition* | | | | | | |
| Pre-Perceived Recognition | **0.42** | 0.03 | 13.15 | <0.001 | 0.36 | 0.49 |
| Non-passing C | **-0.20** | 0.04 | -5.51 | <0.001 | -0.27 | -0.13 |
| Passing C | **-0.22** | 0.03 | -7.31 | <0.001 | -0.28 | -0.16 |
| B | **-0.13** | 0.03 | -4.81 | <0.001 | -0.19 | -0.08 |
| *Predicting Post Identity* | | | | | | |
| Post-Perceived Recognition | **0.92** | 0.02 | 59.02 | <0.001 | 0.89 | 0.95 |
| Pre-Identity | **0.15** | 0.02 | 8.98 | <0.001 | 0.12 | 0.19 |
| Non-passing C | -0.00 | 0.03 | -0.11 | 0.92 | -0.05 | 0.05 |
| Passing C | -0.02 | 0.02 | -1.04 | 0.30 | -0.07 | 0.02 |
| B | -0.00 | 0.02 | -0.03 | 0.98 | -0.04 | 0.04 |

**RQ3: Does gender moderate the relationships between course grade and changes in identity overall and at particular points in the mediated relationship?**

The results of the moderated mediation analysis are shown in Table 6. The indirect effect of grade group on identity through perceived recognition was significantly moderated by gender, with women experiencing stronger negative indirect effects than men. In particular, the moderation happened at the first step in the mediation pathway: women were less likely to feel recognized as a physics person when they received a Passing C or a B. Note that this is on top of the main effects of grade group on perceived recognition. That is, both men and women showed a decline in perceived recognition with lower grades, but the effect was twice as large for women. In contrast, neither the direct effect of grade group on identity nor the relationship between perceived recognition and identity varied by gender.

*Table 6. Unstandardized β, bootstrap standard error, z- score, p-value, and confidence intervals for the relationship between grade group and post identity being mediated by post perceived recognition and moderated by gender, controlling for pre identity and pre perceived recognition. Statistically significant βs are bolded.*

| | Unstandardized *β* | Bootstrap *SE* | *z* | *p* | 95% CI | |
|---|---|---|---|---|---|---|
| *Outcome: Post perceived recognition* | | | | | | |
| Gender | 0.00 | 0.04 | 0.07 | 0.94 | -0.08 | 0.09 |
| Non-passing C | **-0.15** | 0.04 | -3.53 | <0.001 | -0.24 | -0.07 |
| Passing C | **-0.16** | 0.04 | -4.34 | <0.001 | -0.23 | -0.09 |
| B | -0.06 | 0.03 | -1.69 | 0.09 | -0.12 | 0.01 |
| Non-passing C × Gender | -0.11 | 0.07 | -1.49 | 0.14 | -0.25 | 0.03 |
| Passing C × Gender | **-0.15** | 0.06 | -2.54 | 0.01 | -0.27 | -0.03 |
| B × Gender | **-0.21** | 0.06 | -3.77 | <0.001 | -0.32 | -0.10 |
| *Outcome: Post identity* | | | | | | |
| Gender | -0.05 | 0.04 | -1.52 | 0.13 | -0.12 | 0.02 |
| Post perceived recognition | **0.93** | 0.02 | 45.3 | <0.001 | 0.89 | 0.97 |
| Non-passing C | -0.01 | 0.03 | -0.20 | 0.84 | -0.07 | 0.06 |
| Passing C | -0.04 | 0.03 | -1.47 | 0.14 | -0.09 | 0.01 |
| B | 0.00 | 0.03 | 0.02 | 0.99 | -0.05 | 0.05 |
| Non-passing C × Gender | 0.02 | 0.05 | 0.34 | 0.74 | -0.09 | 0.12 |
| Passing C × Gender | 0.05 | 0.05 | 1.05 | 0.29 | -0.04 | 0.14 |
| B × Gender | -0.00 | 0.04 | -0.04 | 0.97 | -0.09 | 0.08 |
| Post perceived recognition × Gender | -0.04 | 0.03 | -1.22 | 0.22 | -0.09 | 0.02 |

The statistically significant pathways are visually presented in Figure 5, highlighting that the effects of gender on the change pathways are restricted to the first mediation step. It also highlights differences between which grade group was significant overall or significantly moderated. To better understand these nuanced differences, we turn to visualizing mean values by grade group and gender at pre and post.

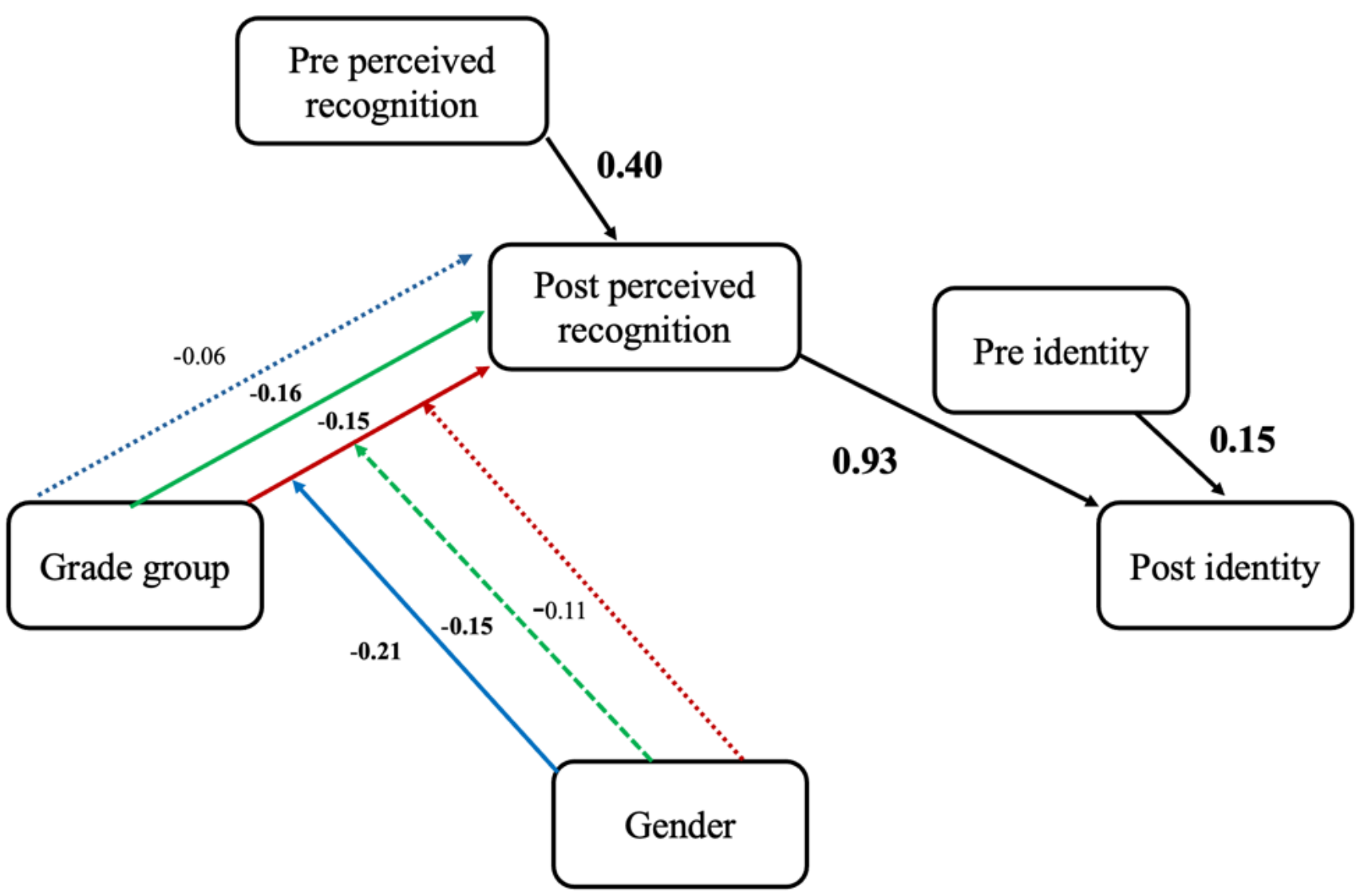

*Figure 5. Statistically significant relationships in the moderated mediation analysis, in which gender moderates the mediation of grades (blue for B, green for passing C, and red for non-passing with respect to A) on identity via perceived recognition. Solid lines represent p<0.001, dashed lines represent p < 0.05, and dotted lines represent p > 0.05.*

Figure 6 presents the mean levels for identity and perceived recognition for men and women at pre and post. At pre, there are moderate-sized gender differences in both identity and perceived recognition, which are largest among those who will receive higher grades. Here, we see that pre-levels are associated with later grades among men but not among women. At post, we see that men and women with As show slight increments in their identity and perceived recognition, and the gender difference effect size declines relative to pre. However, women with grades lower than an A show large declines in their identity and perceived recognition, whereas men show no declines with a B and more muted declines with lower grades. As a result, gender differences at the lower grades become much larger. We also observe that identity and perceived recognition for women receiving a B are lower than those for men receiving a non-passing grade at both pre and post.

Figure A3 and A4 in the Appendix show post identity and post perceived recognition for men and women, controlling for pre identity and pre perceived recognition, respectively.

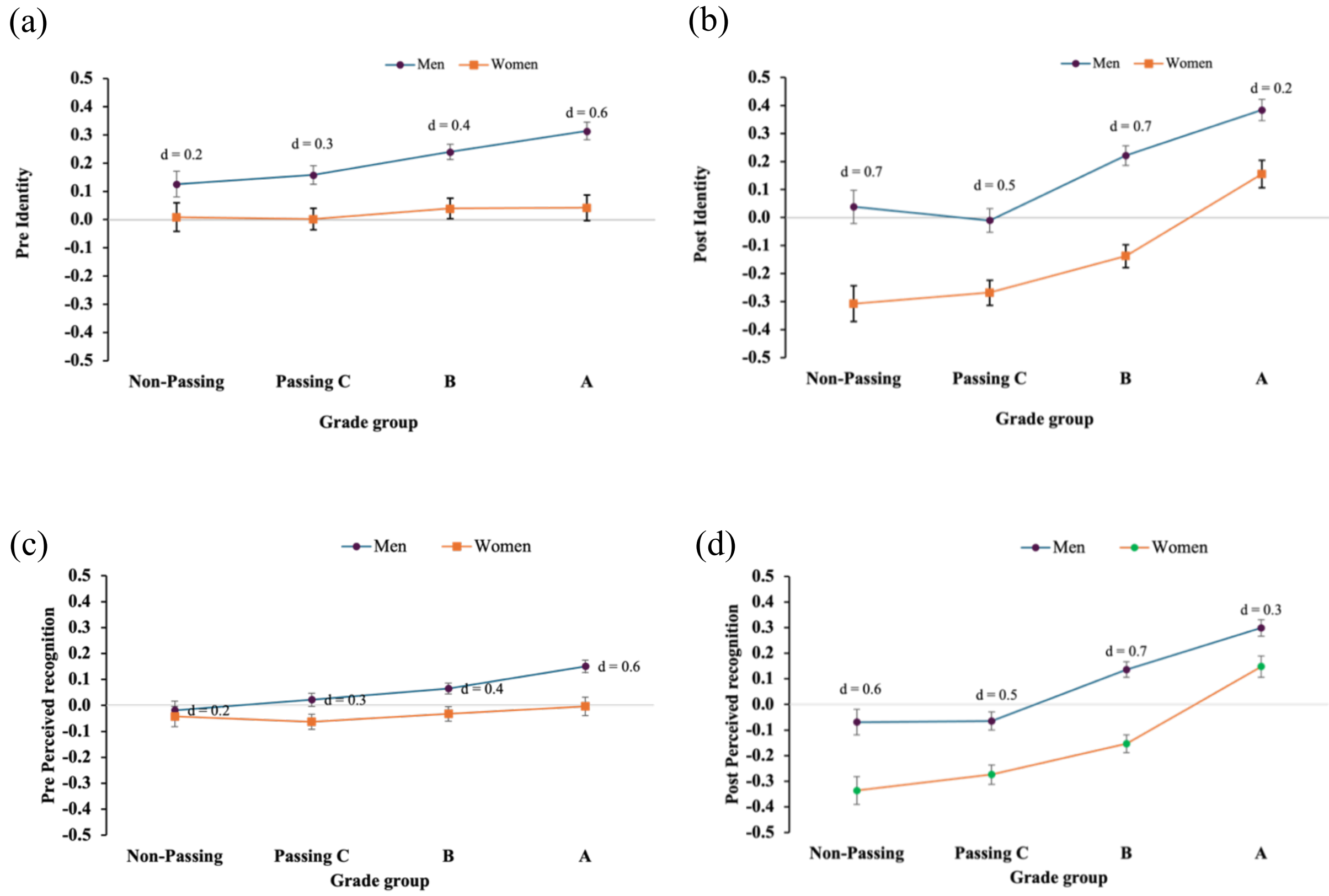

*Figure 6. Predicted means (with SE) for a) pre identity, b) post-identity, c) pre perceived recognition, and d) post-perceived recognition across different grade groups for men and women, along with effect sizes (Cohen's d).*

## Discussion

### Changing Identities and Perceived Recognition as a Function of Course Grades

While prior research has generally supported the theorized relationship between performance and identity (Bottomley et al., 2022; Hazari et al., 2020), the specific relationship with grades has not yet been explored, nor have the relative relationships between grades and perceived recognition or identity been examined. There was a strong relationship between grades and changes in identity.

Going beyond confirming the general predicted relationship, the current study adds important details about this relationship to performance information found in grades. First, perceived recognition was also related to performance. Second, only students receiving an A tended to show growth in their physics identity and perceived recognition. Third, since getting A is relatively rare in this context, this pattern could explain the general declines previously observed in this sequence of introductory courses (Cwik & Singh, 2021a; Li & Singh, 2022; Maries et al., 2021; K. M. Whitcomb et al., 2023). Third, the changes associated with Passing C were similar to or even more negative than the changes found with Non-Passing grades.

The non-linearity of effects might reflect beliefs about student performance and recognition within future, more challenging coursework or career contexts, or they may reflect external attributions students may be more likely to make in the case of very poor performance (e.g., mental health, relative time on other activities, unfair grading practices). Future research is needed to better understand this nonlinearity. Further, additional work is needed to uncover ways of conveying feedback about performance that result in changes in studying and task completion strategies but do not threaten one's identity.

Note that these patterns between grades and identity controlled for the (relatively small) relationship between pre-identity/perceived recognition and grades received. That is, the resulting perceived recognition/identity appears to be driven by grades more than the grades were driven by initial perceived recognition and identity. However, this pattern is consistent with a feedback loop between student beliefs and performance that increases differences among groups (Dorph et al., 2016). Those with higher initial beliefs, like those having higher physics identity or perceived recognition, on average, end up with higher grades and then maintain or even increase their beliefs. By contrast, those with lower initial beliefs, like those having lower physics identity or perceived recognition, on average, end up with lower grades and then further decline their beliefs going into the next course and potentially drop from this physics pathway. The existence of such feedback loops makes early intervention especially important.

**Perceiving Recognition as a Mediator of Performance Effects on Identity**

Prior models have proposed different relationships between performance and identity. For example, many authors have proposed that performance influences identity indirectly through changes in perceived recognition and interest (Godwin et al., 2016; Robynne M Lock et al., 2019). By contrast, (Hazari et al., 2020) argued that performance has both direct effects on identity as well as indirect effects via perceived recognition. However, they focused on perceptions of performance and competence rather than objective performance measures like grades. Here, we show that perceived recognition serves as a key mediator of the relationship with grades in particular, with no direct remaining relationship between grades and identity, and no need to invoke a second pathway via interest. This mediated pathway may provide insights into student sensemaking (Speirs et al., 2026) about grades and possible pathways for intervention. On the one hand, students' reactions to grades seem deeply social, tied to their beliefs about how others respond to grades. Future research is needed to understand whether, in fact, others are changing how they recognize students or whether these are primarily presumed changes. Note that examining the different sources of perceived recognition showed a consistent pattern with grades across sources within the university (instructor, TA), outside the university (family), and those that might span both (friends; see Figure A5 in the Appendix). On the other hand, if instructors and TAs send other messages about recognizing students as belonging in physics, those messages might counteract the negative effects of lower grades on identity and eventual persistence in physics.

## Gender Differences in Responses to Grades

Prior research has found moderate gender differences in students' overall identity and perceived recognition (Cwik & Singh, 2022a; Li & Singh, 2023; Marshman, Kalender, Schunn, et al., 2018; T. J. Nokes-Malach et al., 2019). Importantly, building on these findings, this study shows growing gender differences in identity and perceived recognition across the semester, but particularly in the context of lower grade performance. For example, compared to men, women who get B grade have significantly lower identity and perceived recognition at the end of physics course, and this difference was prominent even controlling for these values at the beginning of the course. Furthermore, the moderating effect of gender on the relationship between grades and changing identities was localized to one particular step: the relationship of grades to perceived recognition. That is, women did not differ in how they reacted to changes in perceived recognition, but rather they appeared to see greater differences in perceived recognition in response to lower grades. Again, we note that future research must examine whether these differences were in assumed changes or whether overt changes in being recognized occurred. It is possible that stereotypes, micro-aggressions, or other "chilly climate" factors shaped how students react to grades (Bottomley et al., 2022). Additional research is needed to unpack the ways in which prior beliefs of students shaped these changes and the extent to which differences in course experiences (such as micro-aggressions from peers after learning about grade performance) played a role.

## Limitations

Given the correlational study design, these findings cannot be taken as conclusive evidence of a causal relation between grades and changes in identity and perceived recognition. However, the study examined changes over time and controlled for prior beliefs, rather than focusing solely on concurrent correlations without controls. Randomly assigning final course grades would not be ethical. However, studies could explore the effects of grade feedback on student identities by counterbalancing stringent and less stringent grading curves across different midterm exams. Researchers could also explore changing message framings alongside grades (Pulfrey et al., 2013), such as providing encouraging comments promoting growth mindset along with the grades. If those changes moderated the current effects, that would not only produce practical guidance for how to mitigate these effects, but it would also support a causal interpretation of the current findings.

This study was conducted at a large research university in a high-enrollment classroom, which provided a sufficiently large sample to ensure adequate statistical power for mediation and moderation analyses. However, this context also limits the generalizability of the findings to smaller, more teaching-focused institutions (Kanim & Cid, 2020), where classroom dynamics and cultural contexts in physics education may differ significantly (Tai & Sadler, 2001). Additionally, students at different universities may vary in their prior preparation and experiences, which can influence their identity development and perceived recognition (Hazari et al., 2010). Future research could also extend this work by examining physics identity and recognition in algebra-based physics courses or in physics courses designed for non-science majors, to better understand how these constructs function across diverse institutional and student populations.

**Pedagogical Implications**

Since identities shape future career pathways and students could persevere to become successful, it is important to foster a learning environment (Miyake et al., 2010; Walton et al., 2015) where students, regardless of their final grades (Starr et al., 2020), can equally see themselves as ‘physics people’. Since women with lower grades tend to report lower physics identity and perceived recognition compared to men, it could be useful to provide targeted support that can help strengthen their sense of identity and increase their perception of being recognized in the physics community.

**Acknowledgement**

Not applicable

**Author Contributions:** Conceptualization, C.D.S. and C.S.; methodology, C.D.S. and C.S.; validation, J.S.K., C.D.S., and C.S.; formal analysis, J.S.K.; investigation, J.S.K., C.D.S., and C.S.; data curation J.S.K., C.D.S.; writing—original draft preparation, J.S.K., C.D.S. and, C.S.; writing—review and editing, J.S.K., C.D.S., and C.S.; visualization, J.S.K.; supervision, C.D.S. and C.S.; project administration, C.D.S. and C.S. All authors have read and agreed to the published version of the manuscript.

**Declaration of interest statement**

No potential conflict of interest.

**Data availability statement**

The datasets used and analyzed during the current study are not available due to data privacy requirements of US FERPA regulations. The data presented in this study are available upon reasonable request from the corresponding author.

**Ethics statement**

The research was approved as exempt from requiring informed consent by the university's Institutional Review Board, given the use of surveys focused on existing educational practice.

# Appendix

## Analysis for checking the consistency of changing Likert scale from 4 point to 7 point

To validate that responses collected using different Likert scale formats could be meaningfully compared, we took advantage of the year in which the survey scale changed. During that year, students completed the pre and post surveys in Physics 1 using the 4-point scale in Fall semester, while students in the spring semester completed the same survey using the 7-point scale in Physics 2. These two sets of survey responses (post in Physics 1 and pre in Physics 2) were collected approximately one month apart. This allowed us to compare the consistency of responses across the two scale formats by examining the relationship between post-survey responses in Physics 1 (using the 4-point scale) and the pre-survey responses of Physics 2 (using the 7-point scale). The correlations observed were both high and stable with for identity and for perceived recognition, suggesting that the scale change did not substantially impact response patterns. In the following two years, the same 7-point scale was used across both semesters. We again examined the correlation between post-survey responses in Physics 1 and pre-survey responses in Physics 2 and found similarly high correlation values, consistent with those observed in the first year. Based on this evidence of stability across formats and years, we proceeded to rescale all survey responses to a common metric and combined the survey results across all three years for subsequent analysis.

**Post identity and perceived recognition, without controlling for pre identity and perceived recognition across different grade groups**

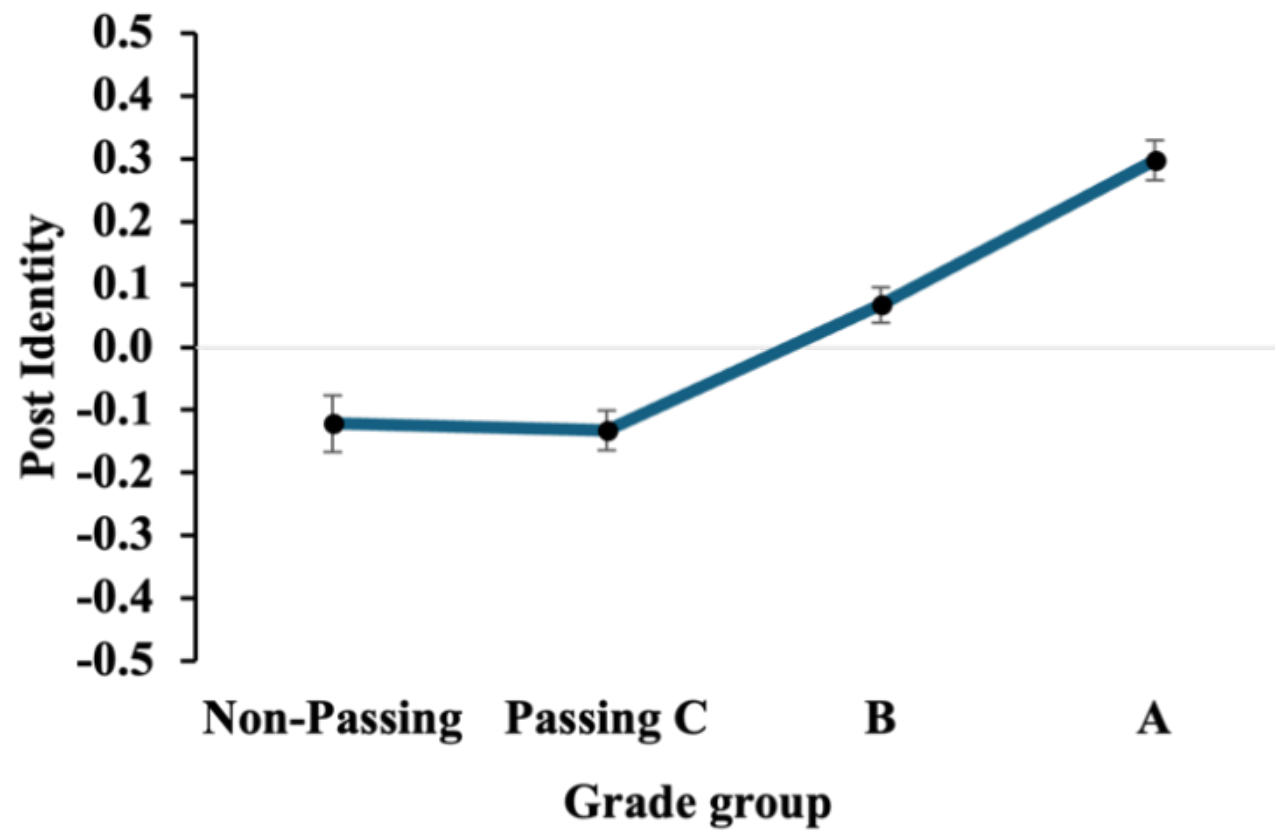

*Figure A1. Predicted means (with* SE*) for post identity across different grade groups, without controlling for pre identity.*

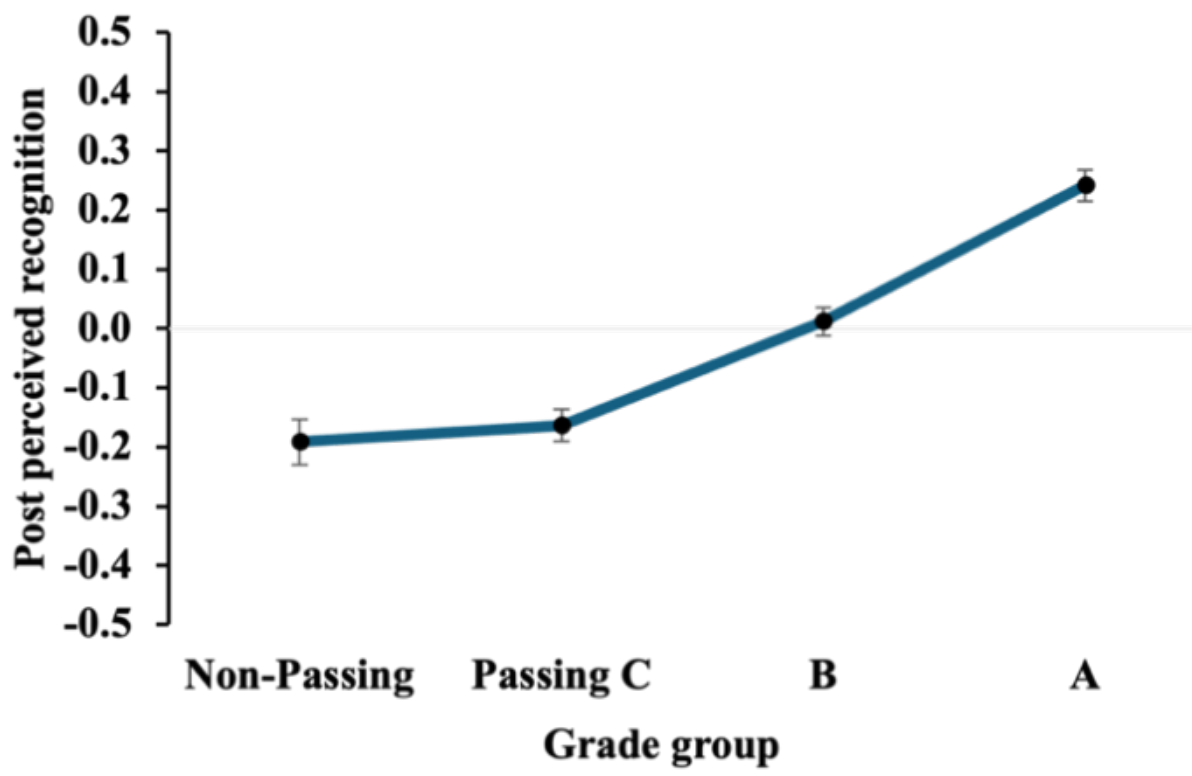

*Figure A2. Predicted means (with SE) for post perceived recognition across different grade groups, without controlling for pre perceived recognition.*

**Post identity and perceived recognition, controlling for pre identity and perceived recognition respectively, across different grade groups for men and women**

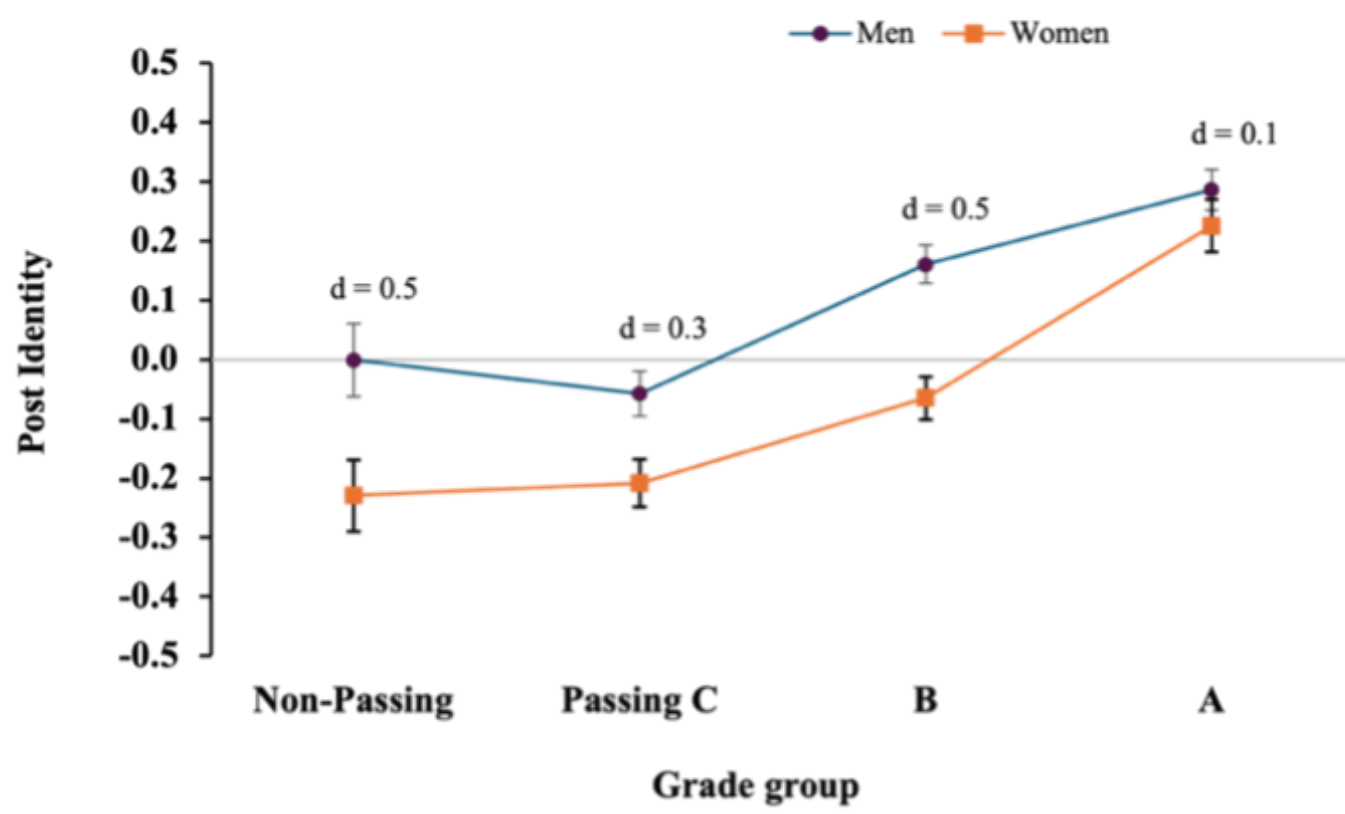

*Figure A3. Predicted means (with SE) for post identity across different grade groups, controlling for pre identity for men and women along with effect size (Cohen's d).*

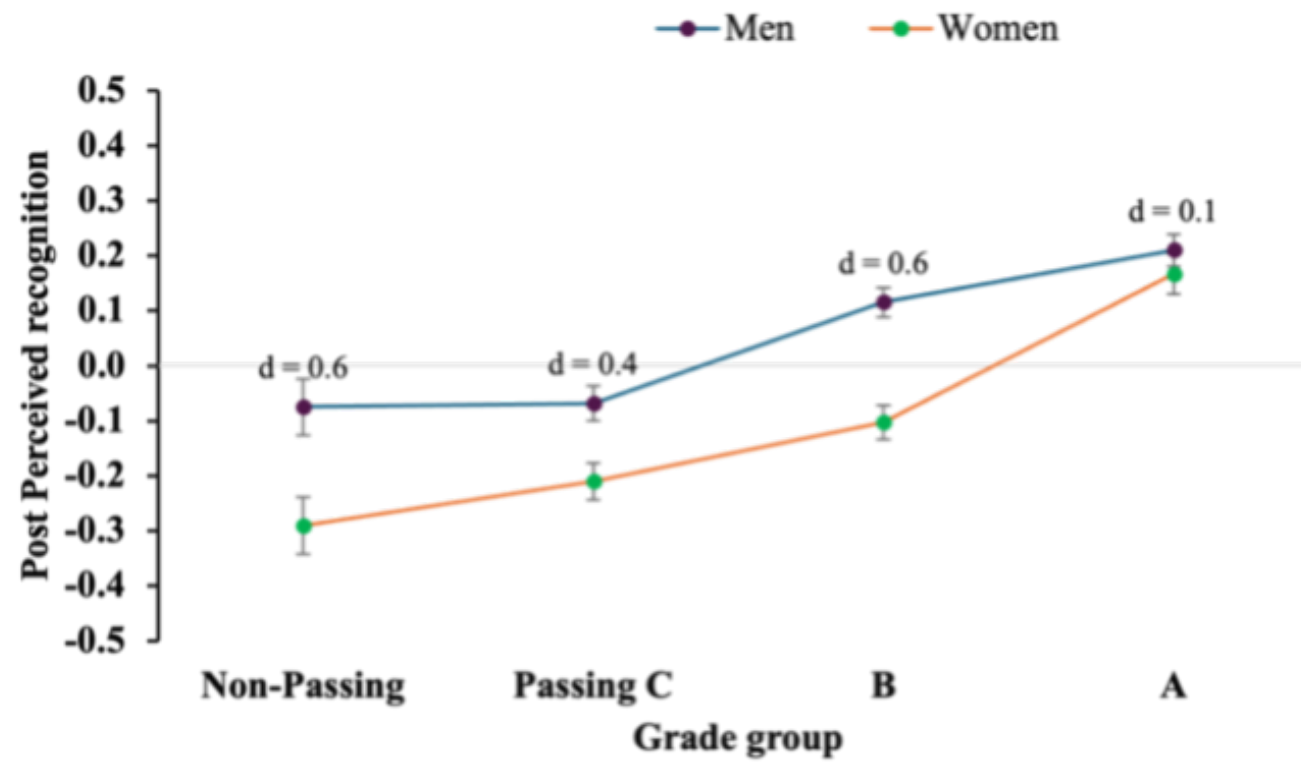

*Figure A4. Predicted means (with SE) for post perceived recognition across different grade groups, controlling for pre perceived recognition for men and women along with effect size (Cohen's d).*

**Post perceived recognition by family, friends, instructor, and TA across different grade groups for men and women, controlling for pre perceived recognition from their respective sources**

The analyses of post perceived recognition by gender and grades were also carried out for each particular source of perceived recognition, and the predicted means controlling for pre levels are presented in Figure  A along with the regression tables A1–A4. The figure shows that the pattern in the overall trend of post perceived recognition was the same across different sources. This suggests that students may hold a consistent internal perception of how others view them, regardless of the specific sources of recognition. These models predict 45%, 42%, 28%, and 28% for family, friends, TA and instructor, respectively. for each source are also shown here.

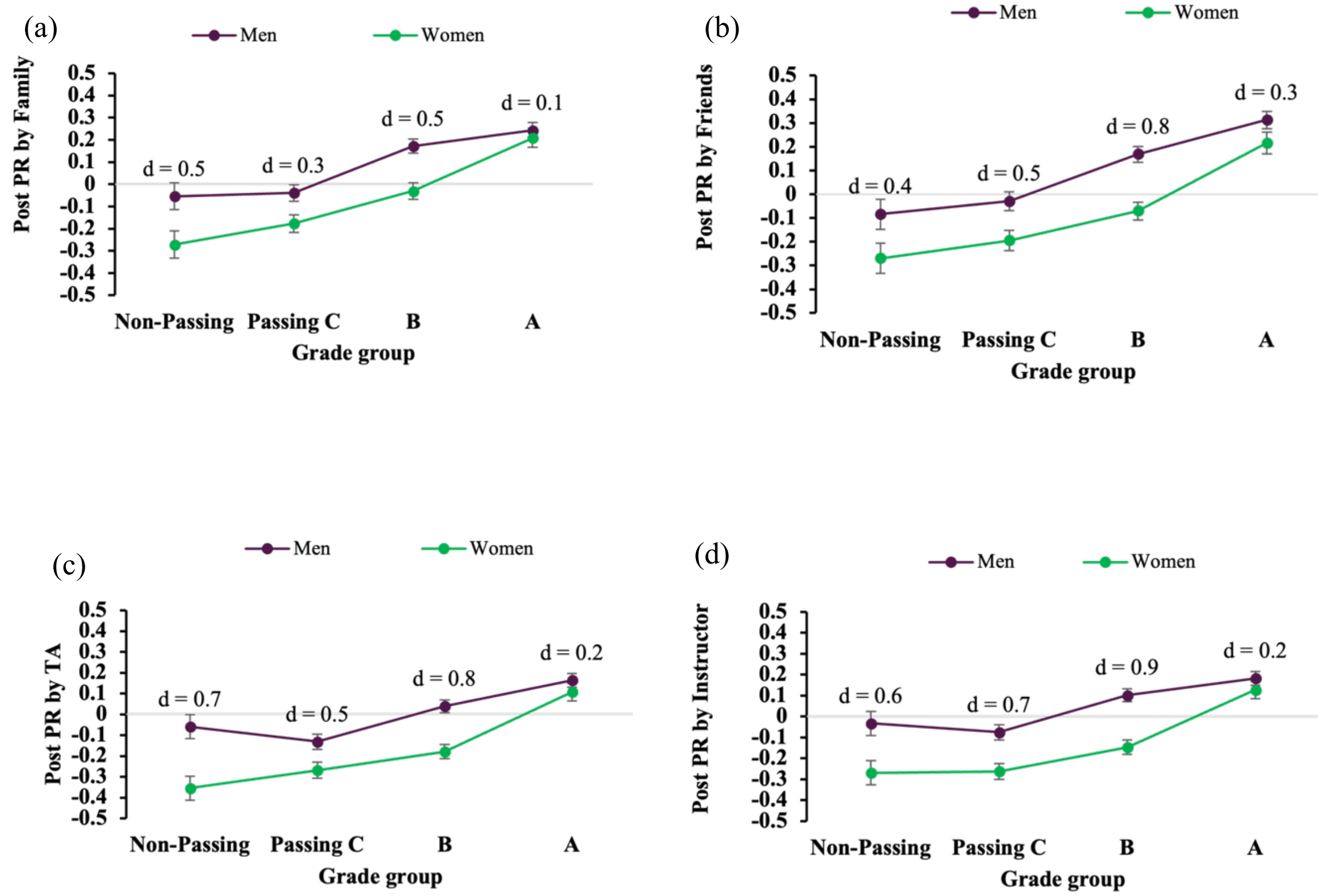

*Figure A5. Predicted means (with SE) for post perceived recognition (controlling for pre preceived recognition) for each item across different grade groups, for men and women after controlling for pre perceived recognition from a) family, b) friends, c) TA and d) instructor along with effect size (Cohen's d).*

*Table A1. Regression coefficients for post perceived recognition by family in relationship to pre perceived recognition, gender, grade group and interaction of grade group and gender, with standardized β, t and p-values.*

| Predictor | Standardized *Beta* | *t* | *p* |
|---|---|---|---|
| Pre perceived recognition by family | 0.57 | 23.43 | <0.001 |
| Women | -0.03 | -0.61 | 0.54 |
| Non-passing | -0.16 | -4.24 | <0.001 |
| Passing C | -0.21 | -5.51 | <0.001 |
| B | -0.06 | -1.52 | 0.13 |
| Non-passing × women | -0.07 | -1.79 | 0.07 |
| Passing C × women | -0.06 | -1.33 | 0.18 |
| B × women | -0.10 | -2.29 | 0.02 |

*Table A2. Regression coefficients for post perceived recognition by friends in relationship to pre perceived recognition, gender, grade group and interaction of grade group and gender, with standardized β, t and p-values.*

| Predictor | Standardized $\beta$ | $t$ | $p$ |
|---|---|---|---|
| Pre perceived recognition by friends | 0.53 | 21.39 | <0.001 |
| Women | -0.08 | -1.68 | 0.09 |
| Non-passing | -0.20 | -5.40 | <0.001 |
| Passing C | -0.25 | -6.41 | <0.001 |
| B | -0.11 | -2.96 | 0.003 |
| Non-passing × women | -0.03 | -0.83 | 0.41 |
| Passing C × women | -0.04 | -0.84 | 0.40 |
| B × women | -0.08 | -1.86 | 0.06 |

*Table A3. Regression coefficients for post perceived recognition by TA in relationship to pre perceived recognition, gender, grade group and interaction of grade group and gender, with standardized β, t and p-values.*

| Predictor | Standardized $\beta$ | $t$ | $p$ |
|---|---|---|---|
| Pre perceived recognition by TA | 0.38 | 13.54 | <0.001 |
| Women | -0.06 | -1.05 | 0.30 |
| Non-passing | -0.14 | -3.37 | 0.001 |
| Passing C | -0.27 | -6.03 | <0.001 |
| B | -0.12 | -2.77 | 0.006 |
| Non-passing × women | -0.11 | -2.45 | 0.02 |
| Passing C × women | -0.05 | -1.08 | 0.28 |
| B × women | -0.12 | -2.30 | 0.02 |

*Table A4. Regression coefficients for post perceived recognition by instructor in relationship to pre perceived recognition, gender, grade group and interaction of grade group and gender, with standardized β, t and p-values.*

| Predictor | Standardized $\beta$ | $t$ | $p$ |
|---|---|---|---|
| Pre perceived recognition by instructor | 0.39 | 14.17 | <0.001 |
| Women | -0.06 | -1.05 | 0.29 |
| Non-passing | -0.14 | -3.30 | 0.001 |
| Passing C | -0.23 | -5.34 | <0.001 |
| B | -0.08 | -1.82 | 0.07 |
| Non-passing × women | -0.08 | -1.88 | 0.06 |
| Passing C × women | -0.09 | -1.79 | 0.07 |
| B × women | -0.14 | -2.79 | 0.005 |